\documentclass{aa}
\usepackage{amsmath,graphics,amssymb}
\usepackage{txfonts}

\allowdisplaybreaks

\newcommand{\dmdt}{\mdot}

\newcommand{\vr}{{v}}

\newcommand{\io}[1]{{#1}_{\text{i}}}
\newcommand{\vo}[1]{{#1}_{\text{H}}}
\newcommand{\he}[1]{{#1}_{\alpha}}
\newcommand\vohe[1]{{#1}_{\text{H}\alpha}}
\newcommand\hevo[1]{{#1}_{\alpha\text{H}}}
\newcommand\voio[1]{{#1}_{\text{iH}}}
\newcommand\heio[1]{{#1}_{\text{i}\alpha}}
\newcommand{\zav}[1]{\left(#1\right)}

\newcommand{\kms}{\,\mathrm{km}\,\mathrm{s}^{-1}}

\newcommand{\de}{\text{d}}
\newcommand{\Teff}{\mbox{$T_\mathrm{eff}$}}

\newcommand{\mdot}{\ensuremath{\dot M}}
\newcommand{\rsun}{\ensuremath{\mathrm{R}_\odot}}
\newcommand{\msun}{\ensuremath{\mathrm{M}_\odot}}
\newcommand\smrok{\ensuremath{\msun\,\mathrm{yr}^{-1}}}

\newcommand\sigorie{$\sigma$~Ori~E}
\newcommand\Gch{\ensuremath{G_\text{Ch}}}

\begin{document}

\title{Multicomponent radiatively driven stellar winds}
\subtitle{IV. On the helium decoupling in the wind of
$\mathbf{\sigma}$
Ori E}
\titlerunning{Multicomponent radiatively driven stellar winds IV.}

\author{J.  Krti\v{c}ka\inst{1} \and J. Kub\'at\inst{2} \and D. Groote\inst{3}}
\authorrunning{J. Krti\v{c}ka, J. Kub\'at, and D. Groote}

\offprints{J. Krti\v{c}ka,\\  \email{krticka@physics.muni.cz}}

\institute{\'Ustav teoretick\'e fyziky a astrofyziky P\v{r}F MU,
            CZ-611 37 Brno, Czech Republic
           \and
           Astronomick\'y \'ustav, Akademie v\v{e}d \v{C}esk\'e
           republiky, CZ-251 65 Ond\v{r}ejov, Czech Republic
	   \and
	   Hamburger Sternwarte, Gojenbergsweg 112, D-21029 Hamburg,
	   Germany}

\date{Received}

\abstract{We study the possibility of the helium decoupling in the
stellar wind of \sigorie.
To obtain reliable wind parameters 
for
this star we first calculate an NLTE
wind model and 
%
derive wind mass-loss rate and
%
terminal velocity.
Using 
corresponding
force multipliers we study the possibility of 
helium decoupling.
We 
find 
that helium decoupling is not possible for realistic 
values
of helium charge (calculated from NLTE wind models).
Helium decoupling 
seems only possible
for 
a
very low helium charge.
The reason 
for
this behaviour is 
the strong coupling between helium and hydrogen.
We 
also find
that frictional heating 
becomes
important in the outer parts of the
wind of \sigorie\ due to the collisions 
between
some heavier elements and the
passive 
components
 -- hydrogen and helium.
For 
a
metallicity ten times lower than the solar
one
both hydrogen and helium decouple from the metals and may fall
back 
onto
the stellar surface.
However, this does not explain the 
observed
chemical
peculiarity since both these components decouple together
from 
the
absorbing ions.
%
Although we do not include the effects of 
the
magnetic field into our
models, we 
argue
that the presence of
a
magnetic field 
will likely not
significantly modify the derived results because
in 
such case
model equations
describe the motion parallel to the magnetic field.
%

}
\maketitle


\section{Introduction}

The radiative force plays an important role in the atmospheres of
early-type stars.
In 
luminous early-type stars 
it
may
be greater than the gravitational force.
In such a case,
the outer stellar layers can
hardly
be in 
hydrostatic equilibrium.
Consequently, the radiatively driven wind is blowing from
%
these stars (see, e.g., Kudritzki \& Puls \cite{kupul}, Owocki
\cite{owopo}, or Krti\v{c}ka \& Kub\'at \cite{kkpreh} for a review).
For many cooler early-type stars the radiative force is not able to
expel the material into the circumstellar environment, however it is
able to cause elemental drift and, consequently, abundance
stratification (e.g.~Vauclair~\cite{vaupreh},
Michaud \cite{mpoprad}) and 
becomes manifested as
chemical peculiarity.

Hot stars with radiatively driven stellar 
winds
and hot chemically
peculiar stars seem to form two diverse groups of stars.
While the flow in the atmospheres of stars with strong winds would
eliminate any possible abundance stratification and chemical
peculiarity, the atmospheres of chemically peculiar stars are supposed
to be very quiet (likely due to magnetic fields) to enable elemental 
diffusion (e.g. Vauclair \cite{vauche}). 

However, there exists a small group of hot stars for which both effects,
stellar wind and chemical peculiarity, seem to be important.
The
sparsely
populated group of helium rich stars probably exhibits both 
effects.
In some cases, 
chemical peculiarity can be even used as a test of
wind existence
(Landstreet, Dolez, \& Vauclair
\cite{nedolez}, Dworetsky \& Budaj \cite{dwobune}). 
Some of the stars with parameters corresponding to hot chemically
peculiar stars may have also purely metallic 
winds, as proposed by
Babel~(\cite{babelb}).

One of the most enigmatic stars of 
the
group of helium rich stars is
the star denoted as {\sigorie}
or \object{HD 37479}.
To our knowledge, this star is the only
known
Bp star with hydrogen lines in emission.
Its
photometrical
and emission line profile variability can be explained assuming the
circumstellar cloud model (Smith \& Groote \cite{smigro}, Townsend et
al. \cite{towog}).
Similarly to the star \object{HD 37776} (Mikul\'a\v sek et al.
\cite{mik}) this star may exhibit rotational braking probably due to
the magnetically controlled stellar wind (Oksala \& Townsend
\cite{oksala}).

Stellar winds of hot stars are accelerated mainly due to the light
scattering in the lines of 
%
elements
like carbon, nitrogen, oxygen or iron. Because these heavier elements
have much lower density
than the bulk of the flow (composed of hydrogen and helium), hot star
winds have a multicomponent nature
(Castor, Abbott, \& Klein \cite{CAK76},
Springmann \& Pauldrach
\cite{treni}, Cur\'e \cite{curdi}, Babel \cite{babela},
Krti\v{c}ka \& Kub\' at
\cite{kk},
\cite{kkii},
hereafter KKII).
Momentum is transferred from low-density heavier elements to the
high-density component composed of hydrogen and helium due to Coulomb
collisions of charged particles.
The 
corresponding
frictional force depends on
the
velocity difference
between wind components.
For very low velocity differences, lower than the mean thermal velocity
(basically for stars with dense winds),
the flow is well coupled.
%
For higher velocity differences
(basically
for stars with low density winds)
the Coulomb collisions are not
able to efficiently transfer momentum from heavier ions to hydrogen and
helium, and frictional heating becomes important
(KKII)
or wind components may
even decouple (Owocki \& Puls \cite{op},
Krti\v{c}ka  \& Kub\' at \cite{kkiii}).

The chemical peculiarity of helium strong stars is in the literature
explained either by radiative diffusion moderated by the stellar wind
(Vauclair \cite{vauche}, Michaud et al. \cite{mihelpek}) or by helium
decoupling in the stellar wind and its consecutive fall back (Hunger
\& Groote \cite{HuGr}).
To test whether the wind decoupling model is able to provide a reliable
explanation of helium chemical peculiarity we decided to calculate
multicomponent wind models suitable for
the
Bp star \sigorie.

\section{NLTE wind model}

\subsection{
Basic
assumptions}

Since there is no accurate determination of \sigorie\ wind parameters
from observation, we decided first to calculate wind mass-loss rate and
terminal velocity using our NLTE wind models (Krti\v{c}ka \& Kub\'at
\cite{nltei}). 

These models make it possible to solve (generally multicomponent)
hydrodynamical equations for radiatively driven winds and enable
prediction of mass-loss rates and terminal velocities.
The
radiative force
and the radiative heating/cooling term are
calculated with NLTE level populations obtained using
atomic data based on
an
extended set of
%
the
TLUSTY
input
files (Hubeny \cite{tlusty}, Hubeny \& Lanz \cite{hublaj}, Hubeny
\& Lanz \cite{hublad}, Lanz \& Hubeny \cite{lahub}).
The data are based on
the Opacity Project (Seaton \cite{top}, Luo \& Pradhan \cite{top1}, Sawey \&
Berrington, Seaton et al. \cite{topt}, Butler et al. \cite{bumez},
Nahar \& Pradhan \cite{napra}), the Iron Project (Hummer et al.
\cite{zel0}, Bautista \cite{zel6}, Nahar \& Pradhan \cite{zel2},
Zhang \cite{zel1}, Bautista \& Pradhan \cite{zel5}, Zhang \& Pradhan
\cite{zel4}, Chen \& Pradhan \cite{zel3}) and the VALD database (Piskunov et al.
\cite{vald1}, Kupka et al. \cite{vald2}).
%
The most important 
simplification
of our code compared to other models
available in the literature 
(Vink et al. \cite{vikolabis}, Pauldrach et al. \cite{pahole},
Gr\"afener \& Hamann \cite{graham}) is the simplified
treatment of the radiative transfer equation.
We use Sobolev approximation for the solution of the radiative transfer
equation in lines (Sobolev \cite{sobolevprvni}, Rybicki \& Hummer
\cite{rybashumrem}) and
we
neglect UV line blocking
as well as
line overlaps.

For the calculation of wind models we assume spherical symmetry.
Since it is well-known that \sigorie\ has a magnetic field
(Landstreet \& Borra \cite{labor})
that generally breaks down the assumption of spherical symmetry, our models,
at least approximately, describe the flow along individual field lines.
Note that in order to obtain a more realistic description of
the
multicomponent wind in 
presence of 
a stellar
magnetic field, it would be necessary to
solve the hydrodynamical equations along
%
field lines with inclusion of proper projections of
individual forces acting on the wind
or to use the magnetohydrodynamical simulations.
In the case of
a
one-component wind model
the former
was done in a simplified form by
Friend \& MacGregor (\cite{frima}),
MacGregor \& Friend (\cite{mafri}),
and more recently
the latter
using
magnetohydrodynamical
simulations by
ud-Doula
\& Owocki (\cite{udo}).

Stellar parameters appropriate for \sigorie\ (Hunger et al.
\cite{huhegr}) are given in Table~\ref{sigoriepar}.
%
Stellar radiative flux
(lower boundary of the wind solution)
of \sigorie\ is
calculated using
the spherically symmetric NLTE model
atmosphere code described in
Kub\' at (\cite{kub}).

Since we intend to study self-initiation of the chemical peculiarity of
\sigorie, we assume solar helium abundance.
If not explicitly stated, we assume
solar chemical composition also for heavier elements.

\begin{table}[hbt]
\caption{Adopted stellar (Hunger et al. \cite{huhegr}) and calculated
wind parameters of \sigorie.}
\label{sigoriepar}
\begin{tabular}{ll}
\hline
effective temperature \Teff & $22\,500\,$K \\
stellar mass $M$ &
$8.9\,\msun$ \\
stellar radius $R_*$ & 
$5.3\,\rsun$ \\
$\log g$
[CGS]
& 3.94 \\
%
\hline\\[-3.5mm]
mass-loss rate
\mdot
& $2.4\times10^{-9}\,\smrok$ \\
terminal velocity $v_\infty$ & $1460\kms$ \\
\hline
\end{tabular}
\end{table}

\subsection{Calculated NLTE model}

\begin{figure}
\resizebox{\hsize}{!}{\includegraphics{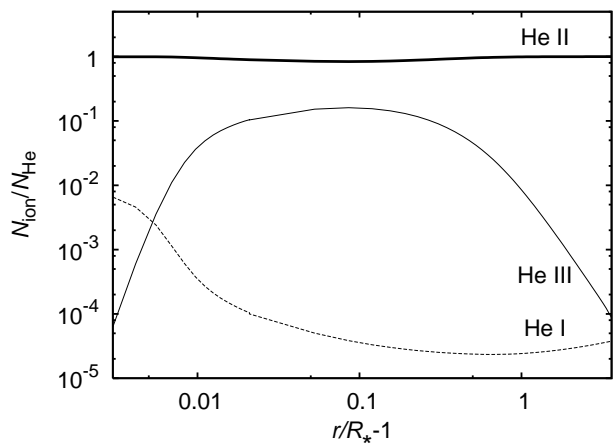}}
\caption{The radial variation of helium ionization fractions
in the wind
calculated
using NLTE model.
\ion{He}{ii}
is the dominant ionization stage
throughout the wind.
Close to the star
(above the photosphere)
the wind density
decreases, consequently the wind becomes more ionized and the ionization
fraction of \ion{He}{i} decreases with the
increasing
radius, while
the fraction
of \ion{He}{iii} increases with the radius.
In the outermost regions the wind becomes slightly
less ionized again due to the decrease of the mean intensity.
}%
\label{helna}
\end{figure}

Using our NLTE wind model,
in which we neglected the multicomponent effects,
we calculated wind parameters for \sigorie.
The
predicted wind mass-loss rate $\dot M=2.4\times10^{-9}\,\smrok$ is in
agreement with order of magnitude estimate inferred from observations
(see the discussion in Groote \& Hunger \cite{grhu}).
The derived terminal velocity $v_\infty=1460\kms$ implies the ratio
of 
terminal velocity to 
escape velocity $v_\infty/v_\text{esc}=1.8$
(note that the radiative force due to 
light scattering on free
electrons has
only
a small influence here since the Eddington parameter
$\Gamma\approx0.02$
is low).
%
The derived ratio of terminal velocity to escape velocity
is consistent with that for stars close to the bistability jump
(see Pauldrach \& Puls \cite{papubista},
Lamers et al. \cite{lsl}, Crowther et al. \cite{crobis}).

Since we 
want
to study helium decoupling that is charge-dependent
(as follows from the frictional term due to the Coulomb collisions
-- see Eq.~\eqref{trekon} or Eq.\,40 in KKII)
we have to study the calculated variation of He ionization in the wind.
From Fig.~\ref{helna} we conclude that helium is mostly singly ionized
in the stellar wind with only a small contribution of doubly ionized
helium.
The fraction of neutral helium
in the wind is very small, less than $10^{-4}$
(for $r\gtrsim1.01R_*$).

\subsection{Multicomponent effects}

Before studying helium decoupling we shall discuss another aspect of a
wind model for {\sigorie}, namely, multicomponent effects due to the
friction of individual heavier elements with hydrogen and helium.

Since the radiative acceleration acting on individual heavier elements
is different, 
each element
%
moves
with slightly different 
velocity.
The dimensionless velocity difference between
given
individual
heavier element $h$
%
and hydrogen and helium
\begin{equation}
\label{xdif}
%
x_{h\text{p}}=\frac{{\vr}_{h}-{\vr}_\text{p}}{\alpha_{h\text{p}}}
\end{equation}
is roughly equal to (Krti\v cka \cite{nlteii}) 
\begin{equation}
\label{xipp}
x_{h\text{p}}\approx
{g}_{h}^{\mathrm{rad}}\frac{m_{h}}{{n}_\text{p}}
\frac{3kT}{8\sqrt\pi  {q}_{h}^2{q}_\text{p}^2\ln\Lambda}.
\end{equation}
We use the subscript $h$ to denote individual heavier
elements -- i.e.~$h$ stands for N, Si, S, etc. For this purpose we
do not use the subscript~i that denotes all heavier elements described
together. The
%
subscript p 
denotes
the passive component (hydrogen and helium).
%
Here
${\vr}_\text{p}$ and ${\vr}_{h}$ are 
the
radial velocities of
the
components,
$m_\text{p}$ and $m_h$ are
their atomic masses,
$q_h$
and ${q}_\text{p}$ their charges,
${g}_h^{\mathrm{rad}}$
is the radiative acceleration acting on given
element~$h$,
${n}_\text{p}$
is
the number density of hydrogen and helium, $T$ 
is
the
temperature (assuming that the temperatures of all wind components are the same), $\ln\Lambda$ 
is
the Coulomb logarithm, and finally,
\begin{equation}
%
\label{alrov}
\alpha_{h\text{p}}^2=\frac{2kT\zav{m_\text{p}+m_h}}{m_\text{p}m_h}.
\end{equation}

\begin{figure}
\resizebox{\hsize}{!}{\includegraphics{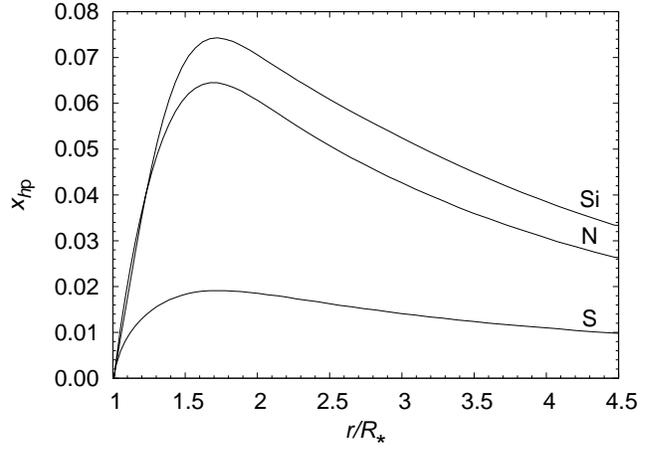}}
\caption{Calculated relative velocity 
differences
%
(see Eq.~\eqref{xdif})
between individual
heavier elements and 
the
passive
component.
Velocity differences for elements with low $x_{h\text{p}}$ values are
not plotted in this graph.
Note that 
here
the effects of frictional heating are neglected.}
\label{rozdil}
\end{figure}

\begin{figure}[htb]
\resizebox{80mm}{!}{\includegraphics{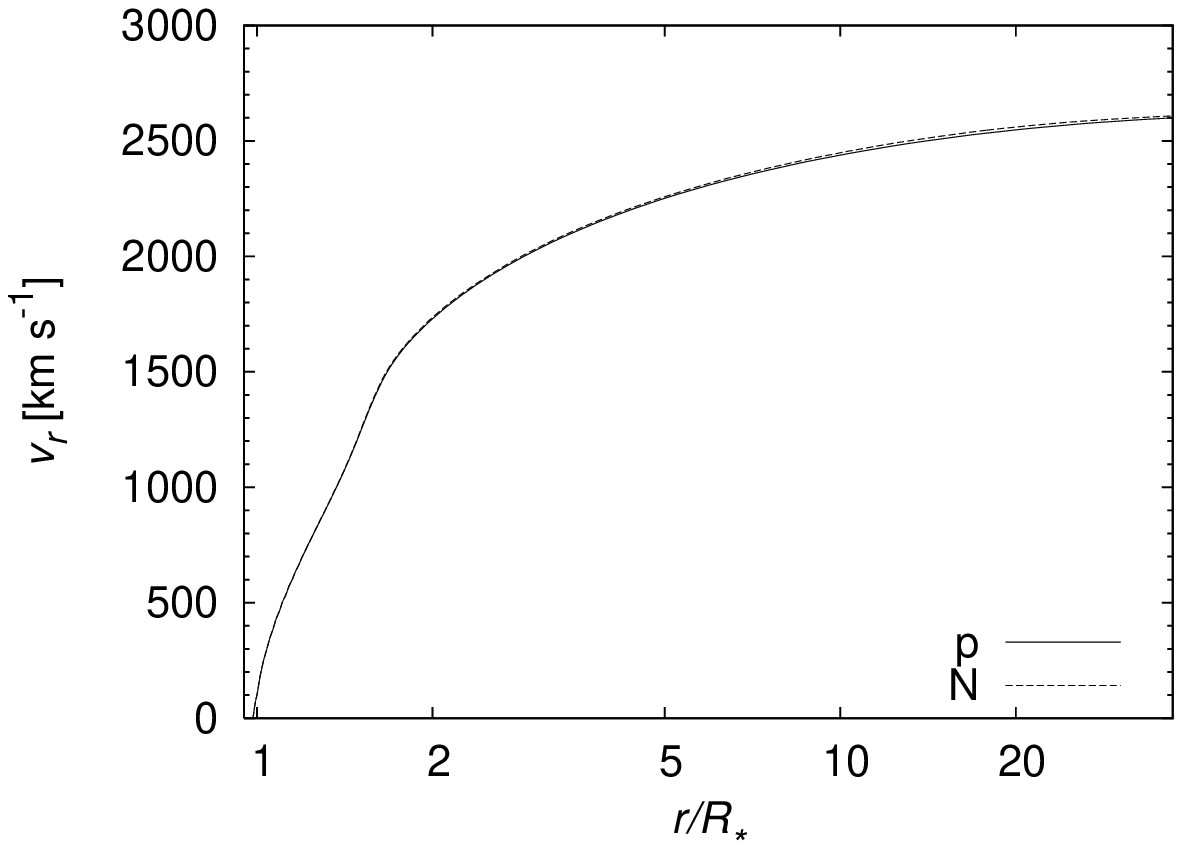}}
\resizebox{80mm}{!}{\includegraphics{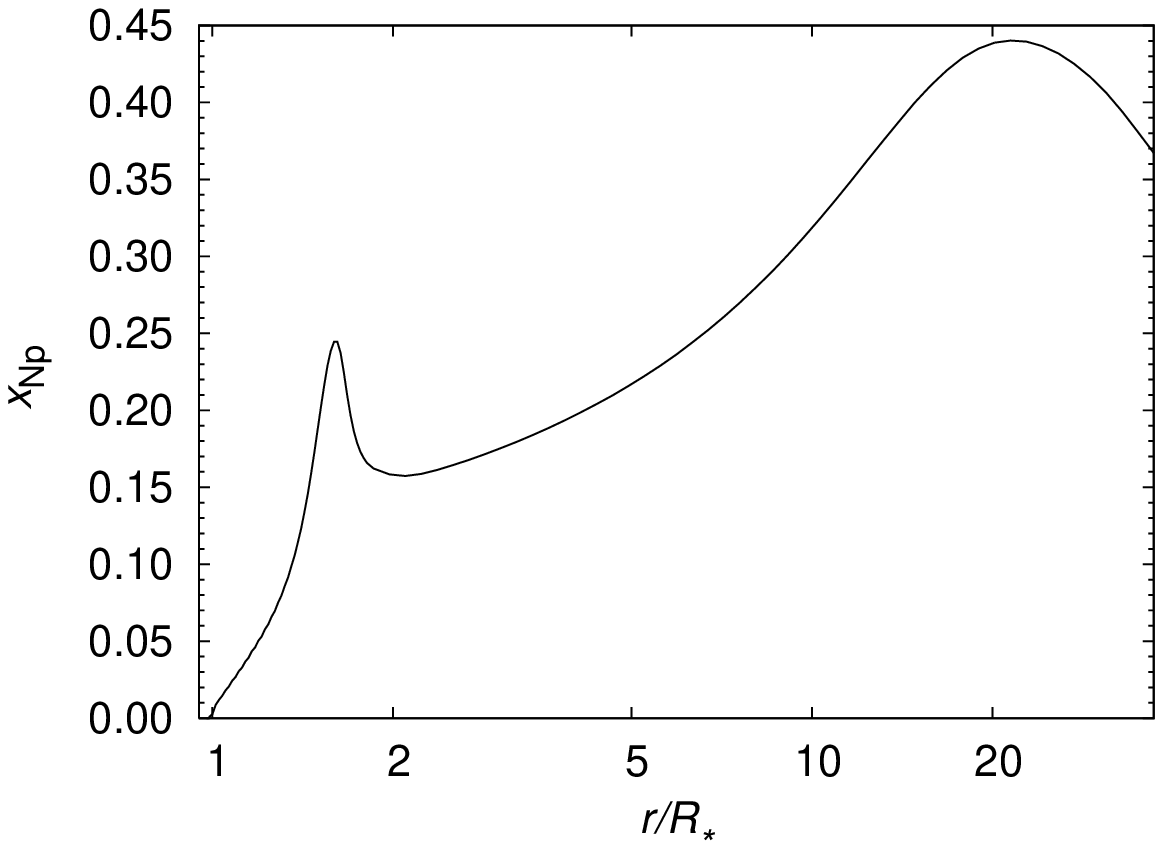}}
\resizebox{80mm}{!}{\includegraphics{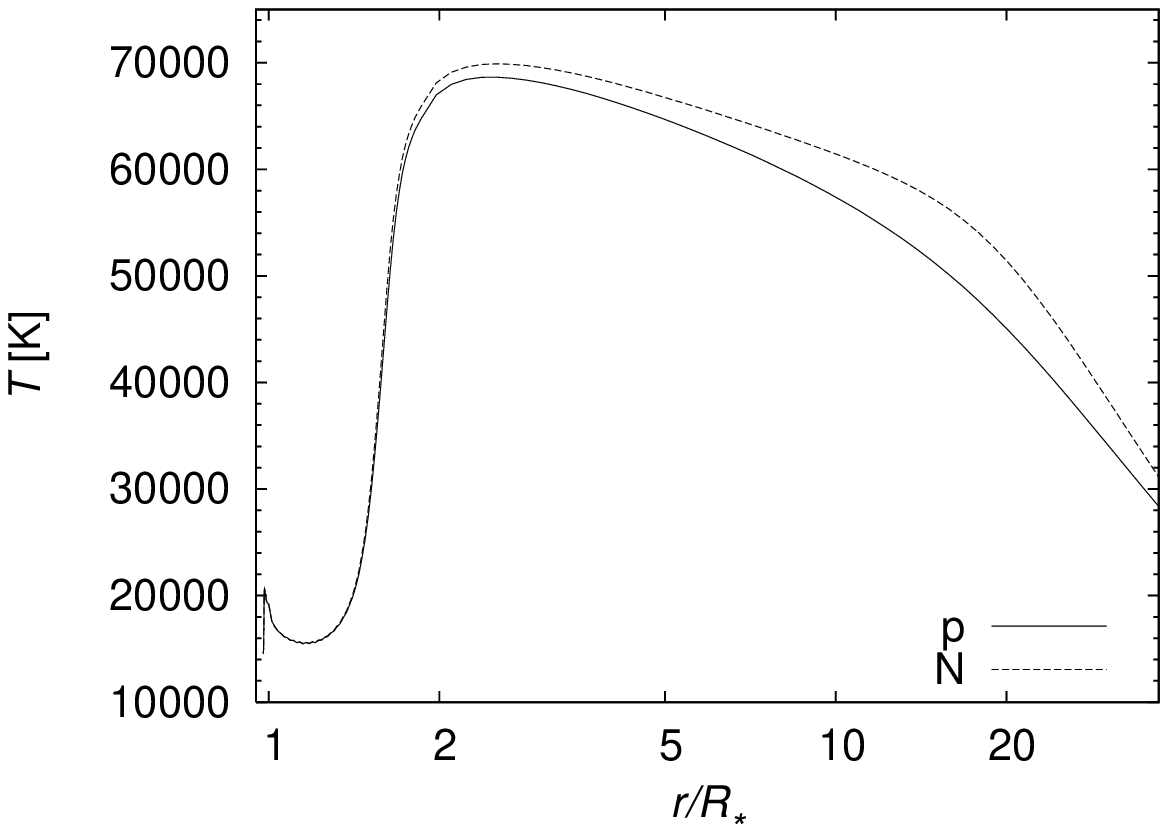}}
\caption
{Four component NLTE wind model with components hydrogen and
helium
(p),
electrons, nitrogen
(N)
and remaining heavier ions.
{\em Upper panel:} Radial velocity of 
the
passive 
component 
p
and nitrogen.
The
velocities of wind components
are nearly 
indistinguishable in the plot.
{\em Middle panel:}
Plot of dimensionless velocity difference
between 
the passive component
and nitrogen
%
$x_\text{Np}$
(Eq.~\ref{xdif}).
%
The velocity 
differences
between wind components 
are
relatively high, enabling frictional heating (as shown in the plot of the
temperatures of individual wind components, {\em lower panel}).}
\label{dusik}
\end{figure}

Using Eq.~(\ref{xipp}) we calculated the approximate velocity 
differences
between hydrogen and helium and
individual
heavier ions (see Fig.~\ref{rozdil}).
Apparently,
the
relative velocity differences between silicon or nitrogen
and the hydrogen-helium component are so high that frictional heating
may 
be
enabled. To test this possibility we calculated 
four-component wind models.
Despite the fact that the relative velocity difference between silicon
and 
the
passive component is slightly higher than that of nitrogen, we
selected 
the
nitrogen component for this test because nitrogen has higher
abundance and, consequently, 
heating effects caused by friction
are expected to be higher
for this element.
Resulting four-component wind model (with wind components hydrogen plus
helium, free electrons, nitrogen and the remaining heavier ions) 
is
given in Fig.~\ref{dusik}.
Apparently, the stellar wind temperature
significantly increases in the outer parts due to 
frictional heating.
However,
no decoupling occurs in this case
(since 
$x_\text{Np}$
is too low to
enable
the decoupling).
The wind terminal velocity significantly increases compared to the case
when the frictional heating is neglected.
This is caused by the change in the ionization and excitation
equilibrium 
due to
a
higher wind temperature.
Since this effect occurs for velocities higher than that
at
the
critical point, the mass-loss
rate is not essentially modified in 
four-component models.
A
similar effect, 
but to a lesser extent, was already reported
by Krti\v cka
(\cite{nlteii}).

Because the effects of frictional heating between heavier elements (in
our case nitrogen) and 
the
passive component are important basically only
for velocities higher than the escape 
velocity
(which is
$v_\text{esc}=790\kms$ at the stellar surface), they do not
significantly modify the flow properties close to the stellar surface.
Since this region is important for the study of the helium decoupling
and its
reaccretion,
we neglect the effect of friction caused by the distinct properties of
all individual heavier elements.
However,
%
heating due to
%
friction between hydrogen and helium ions is properly taken
into account.

\section{The possibility of helium decoupling}

\subsection{Model assumptions}

To test the possibility of helium decoupling we calculated
four-component wind models with 
the
following four explicitly included wind components: hydrogen, helium,
heavier ions and free electrons.
For this purpose we did not use our NLTE models, but 
%
simpler (and faster) 
models with the radiative force in the CAK approximation (Castor, Abbott \&
Klein \cite{cak}; see KKII
for
the description of these models).
The parameterized description of the radiative force in these models enables us
to vary both wind parameters and wind ionization state independently
%
to discuss the parameter space for which helium decoupling seems possible.
%

%
To model mass and charge of heavier ions 
%
we selected parameters corresponding to oxygen
(because elements like C, N, O, Si, S are important for radiative driving).
The selection of oxygen 
does
not have a significant impact on derived results.
%
The CAK force parameters
\begin{align}
\label{cakpar}
k&=0.38,&\alpha&=0.50,&\delta&=0.03
\end{align}
(CAK, Abbott \cite{abpar},
Puls et al.~\cite{pusle})
are chosen to obtain the same mass-loss rate
and 
a
similar velocity profile as 
derived from
%
our
NLTE models.
Note that the terminal velocity calculated using these parameters is
slightly higher than 
calculated
from NLTE models.
However, this does not significantly influence our
analysis.
%


\subsection{Realistic helium charge}

We calculated four-component wind models with the realistic value of the
helium charge taken from previously discussed NLTE 
model.
Since helium is mostly singly ionized in the wind region
(see Fig.~\ref{helna}),
we simply assume $\he z=1$,
%
where
$\he z$ is the relative charge in
units
of the elementary charge $e$, 
$\he q=e\he z$.

\begin{figure}
\resizebox{\hsize}{!}{\includegraphics{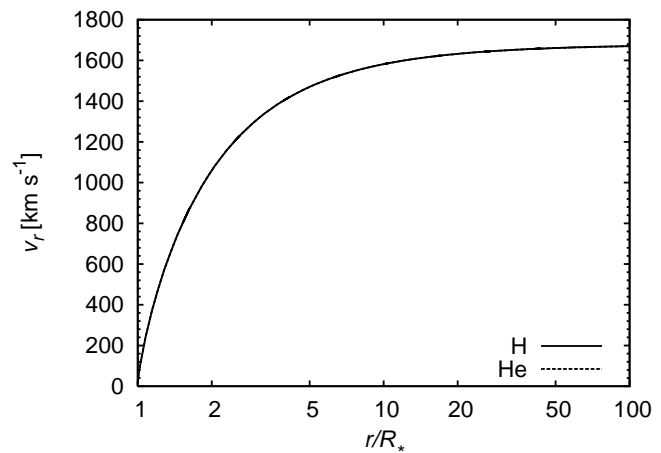}}
\caption{Calculated
velocity structure of
a four-component
(H, He, 
heavier
ions, electrons)
wind model for 
a
realistic value of helium
charge $\he z=1$.
%
Hydrogen and helium velocities are nearly
identical and helium does not decouple.}
\label{z100q100}
\end{figure}

The velocity 
profile
of a calculated four-component wind model is shown in Fig.~\ref{z100q100}.
Evidently,
the
radial velocities of all wind components are nearly the same and the
wind components are well coupled.
Consequently, helium decoupling cannot be found in this case.

\subsection{Low helium charge}

\begin{figure}
\resizebox{\hsize}{!}{\includegraphics{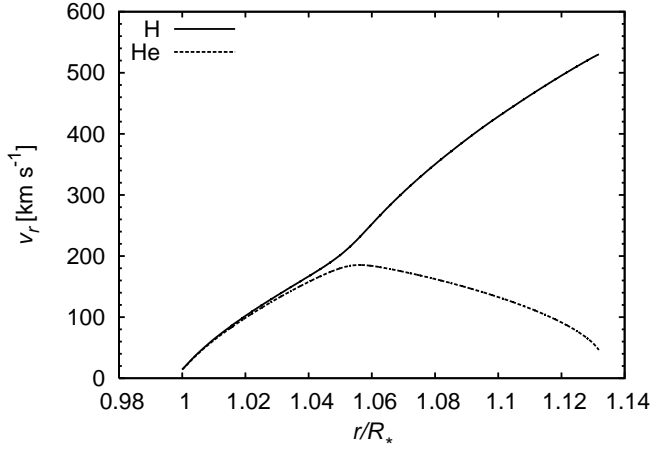}}
\caption{The velocity structure of a calculated four-component wind
model for a very low 
probably non-realistic
value of helium charge $\he z=0.01$.
Helium is accelerated at the wind base, however due
its low charge it subsequently decouples and may fall back.}
\label{z100q001}
\end{figure}

For very low helium charge, significantly lower than that calculated
from NLTE equations, helium decoupling 
is
possible.
As an example we plot 
in Fig.~\ref{z100q001}
a comparable wind model but for helium charge $\he z=0.01$.
Apparently, the wind density at the wind base is high enough to
accelerate helium from the stellar surface to supersonic velocities.
However, due to its  very low charge helium decouples from the mean flow
at velocities lower than the escape velocity.
Consequently, helium remains
bound
to the star and may fall back to the stellar surface.
Hydrodynamical simulations are necessary to study this process in detail
(cf.
Porter \& Skouza \cite{reac}, Votruba et al. \cite{ufosap}).

\subsection{Helium coupling}

The main result of this paper -- that for realistic values of
its
charge helium
does
not decouple from the wind flow for parameters suitable for a star like \sigorie\ -- 
seems
to be in contradiction to the result of Hunger \& Groote (\cite{HuGr})
who argued on the basis of the work of Babel (\cite{babela},
\cite{babelb})
for the decoupling hypothesis that explains 
observational results.
%
To find the reason for this difference we 
study the hydrodynamical
equations analytically.
The equation of motion for helium (neglecting
the
radiative force acting on
helium, 
the
gas-pressure term and 
the
charge separation electric field) has the
form of (Burgers \cite{burgers})
\begin{equation}
\label{rovhyb}
\he\vr\frac{\de \he\vr}{\de r}=
-g+\frac{1}{{\he\rho}} \sum_{a=(\text{H},\text{e},\text{i})}
 K_{\alpha a}
\Gch
(x_{\alpha a})\frac{{\vr}_a-{\he\vr}}{|{\vr}_a-{\he\vr}|},
\end{equation}
where ${\he\vr}$ is the helium radial velocity, ${\he\rho}$ is the
helium mass density, 
%
$
\Gch
(x)$
is the Chandrasekhar function,
and $g$ is the
gravitational
acceleration, $g=GM/r^2$.
The summation of the frictional force in
Eq.~(\ref{rovhyb})
takes into account
friction due to hydrogen "H", 
free
electrons "e" and
all heavier ions "i"
(
the latter are
described as one component).
%
The
%
frictional coefficient $K_{ab}$
due to the Coulomb collisions
between any 
component
$a$ and $b$
is given by
\begin{equation} 
\label{trekon}
{K}_{ab}={n}_a{n}_b\frac{4\pi  {q}_a^2{q}_b^2}{k T_{ab}}\ln\Lambda,
\end{equation} 
where ${n}_a$ and ${n}_b$ are the number densities of individual components
and ${q}_a$, ${q}_b$ are charges of these components.
The mean temperature is
\begin{equation}
\label{tab}
T_{ab}=\frac{m_aT_b+m_bT_a}{m_a+m_b}
.
\end{equation}

In the case of the helium decoupling, helium has to decouple from
\emph{both}
hydrogen and heavier ions.
The maximum possible value of the Chandrasekhar function is in both
cases the same, however, the value of the frictional coefficient is
different,
$K_{\alpha\text{H}}\gg K_{\alpha\text{i}}$ since $n_\text{H}\gg n_\text{i}$
(assuming
an
ionized
medium).
This means that hydrogen-helium collisions are extremely important for
the dynamics of helium component.
%
Consequently,
hydrogen
may be
%
(depending on its ionization state)
much more important for the helium acceleration than 
the
heavier
ions.
%
The neglected strong coupling between hydrogen and helium is the main
reason why Hunger \& Groote (\cite{HuGr}) concluded that the helium
decoupling from flow of hydrogen and passive plasma is possible (even
for 
realistic
parameters of wind plasma
-- like helium ionization state).

However, the situation may differ from the extreme case described above.
The contributions of hydrogen and heavier ions to helium acceleration
may be equally important due to different values of the relative
velocity 
differences
$x_{\alpha\text{H}}$ and $x_{\alpha\text{i}}$ that
are 
arguments of the Chandrasekhar function (see Appendix
\ref{dvprib} for a more detailed approximate description of the flow).
However, in the case when the velocity difference between helium and
heavier ions is large, $x_{\alpha\text{i}}\gtrsim1$, the frictional
force due to hydrogen dominates.
Note that after the decoupling of helium and hydrogen
(in the particular case displayed in Fig.~\ref{z100q001} for $r/R_*\gtrsim1.06$),
helium still
remains coupled to electrons.
This is due to the fact that
$\alpha_{\alpha\text{e}}\gg\alpha_{\alpha\text{H}}$.
The coupling of helium to electrons, however,
modifies the helium velocity only slightly.

The only possibility how to enable helium fall back onto the stellar
surface is to break the strong coupling between helium and hydrogen
while keeping coupling between hydrogen and heavier ions.
The only way how to secure this in our 
self-consistent
models is to lower the helium
charge.
This is why for very low helium charge helium decoupling is possible. 

We may
derive an approximate condition for the helium decoupling in
the stellar wind.
If helium decouples from the stellar wind and subsequently falls back,
there is a point where the helium velocity gradient is zero (see
Fig.~\ref{z100q001}).
The helium momentum equation Eq.~\eqref{rovhyb} has in this point the
form of 
\begin{equation}
\label{oddelbo}
g=\frac{1}{{\he\rho}} \vohe K 
\Gch
(\vohe x)
\end{equation}
(neglecting friction with heavier ions and electrons).
%
The Chandrasekhar function is
limited
by its maximum value 
$
\Gch
(\vohe x)\leq G_\text{max}$.
In the case of decoupling the frictional acceleration on the right hand
side of Eq.~\eqref{oddelbo} is lower than the
gravitational
acceleration
(inequality in \eqref{oddelbo} occurs).
With $g=GM/r^2$, the hydrogen mass-fraction 
$X\approx0.71$,
and using Eq.~\eqref{trekon} the condition 
for
helium decoupling for
the
radial velocity $v$ is
\begin{equation}
\label{heoddelpod}
\frac{\vo m\he m kGM}{e^4X\ln\Lambda G_\text{max}}\gtrsim
\frac{\dot M\vo z^2\he z^2}{vT},
\end{equation}
where $e$ is the elementary charge.
Note that the wind temperature $T$ (or, explicitly, 
the
mean temperature
$\vohe T$, see Eq.\eqref{tab}) is in our case usually influenced by the
frictional heating.
Using scaled quantities,
\eqref{heoddelpod}
can be rewritten as
\begin{equation}
\label{heoddelpodskal}
\he z^2 \zav{\!\frac{\dot M}{10^{-14}\,\text{M}_\odot/\text{year}}\!}\!
\lesssim\frac{3}{\vo z^2}\!
\zav{\!\frac{M}{1\,\text{M}_\odot}\!}\!
\zav{\!\frac{T}{10^5\,\text{K}}\!}\!\zav{\!\frac{v}{100\kms}\!}.
\end{equation}
A
similar condition was obtained by Owocki \& Puls (\cite{op}, Eq.~(23)
therein),
but
for coupling between hydrogen and heavier ions.
%
The
inequality
\eqref{heoddelpodskal}
gives condition 
for the helium charge
that 
enables
helium decoupling.                               
For example, for values taken from 
the
model displayed in
Fig.~\ref{z100q001} we obtain the relative helium charge necessary for
decoupling
%
$\he z\approx0.01$, consistent with our previous detailed
calculations.

\subsection{The possibility of the decoupling of neutral helium}

Our conclusion that helium decoupling is unlikely in the case of
realistic helium charge is based on the assumption that
all helium ionization stages
can be treated as a
single
wind component.
This physically means that the processes of ionization and recombination
are so rapid that the typical time 
of 
recombination or ionization
%
is much
shorter
than the
characteristic time of change
in
flow properties (i.e.~velocity and
density).
While the violation of this condition would not affect the results in
the case of doubly ionized helium (because it has higher charge than the
singly ionized helium and, consequently, is more tightly coupled to
hydrogen, see 
Eq.~\ref{trekon}),
the violation of this condition in the case of neutral
helium may
offer an interesting possibility of a helium decoupling mechanism.
%
The reason for this is that the cross-section of collisions of protons
with neutral helium is much lower than the effective cross-section for
collisions between charged particles (Krsti\'c \& Shultz \cite{voda}).

To test this, we compare the
characteristic time of change
in
flow properties
(roughly
corresponding to the time during which the velocity increases 
to  
its
actual value)
\begin{equation}
\label{tcas}
\tau_\text{flow}=\zav{\frac{\de \vr}{\de r}}^{-1}
\end{equation}
with the typical time of ionization of \ion{He}{i} 
\begin{equation}
\label{tauioniz}
\tau_\text{ioniz}=\frac{1}{P_{\ion{He}{i}\rightarrow\ion{He}{ii}}},
\end{equation}
where $P_{\ion{He}{i}\rightarrow\ion{He}{ii}}$ is 
the
ionization rate of
\ion{He}{i} (Mihalas \cite{mihalas}).
This value was taken from our NLTE models. Since excited levels of
neutral helium have very low occupation numbers, we neglected their
contribution to $P_{\ion{He}{i}\rightarrow\ion{He}{ii}}$.

\begin{figure}
\resizebox{\hsize}{!}{\includegraphics{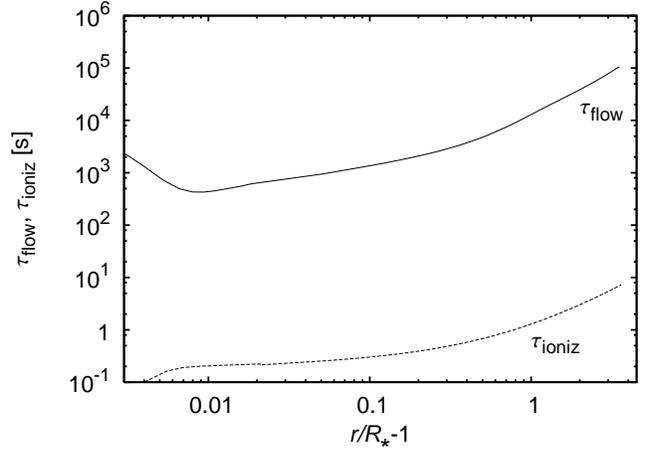}}
\caption{Comparison of the characteristic 
time of change
in
flow properties
$\tau_\text{flow}$
and 
the
characteristic time of \ion{He}{i} ionization
$\tau_\text{ioniz}$}
\label{casy}
\end{figure}

Comparison of characteristic times $\tau_\text{flow}$ and
$\tau_\text{ioniz}$ is given in Fig.~\ref{casy}.
The 
characteristic time $\tau_\text{flow}$
first slightly decreases with radius and then increases again.
This marks the region of fast wind
acceleration.
Also the \ion{He}{i} ionization time
increases  with radius. However, the
characteristic time
$\tau_\text{flow}$ is much larger than the \ion{He}{i} ionization time
$\tau_\text{ioniz}$.
Consequently, helium may be treated as one component
and no decoupling between different helium ionization stages occurs.

\section{Models with lower metallicity}

Since the metallicity of individual surface elements of \sigorie\ varies
with location on the stellar surface,
we studied also the influence of different metallicity on the
possibility of helium decoupling. In particular, we selected the
metallicity to be $Z/Z_\odot=0.1$, as derived by Smith \& Groote
(\cite{smigro}) for \sigorie.

\subsection{
One-component
NLTE wind model}

Our NLTE wind model for metallicity $Z/Z_\odot=0.1$
(not shown here)
predicts a very low
mass-loss rate of $\dot M=7.4\times10^{-11}\,\smrok$. This is much lower
than that can be inferred from simple relation 
$\dmdt\sim Z^{0.67}$
derived by
Krti\v cka (\cite{nlteii}) for
SMC
O stars.
This 
%
indicates that for lower wind densities the dependence of
mass-loss rate on 
metallicity is more pronounced.

Steeper
metallicity dependence of mass-loss rates for small
metallicities may seem unexpected.
%
To explain this behaviour,
let us first discuss the
contribution
of lines of different elements to the
radiative force. 
%
This is connected with the value of
the line optical depth
$\tau$
which is
in the Sobolev approximation 
%
given by (e.g.~Abbott \cite{abpar}, Eq.\,1,
neglecting the finite-disk correction)
\begin{equation}
\label{tausob}
\tau\approx\frac{\pi e^2}{m_\mathrm{e}\nu_{ij}}
\zav{\frac{n_i}{g_i}-\frac{n_j}{g_j}} g_if_{ij}\zav{\frac{\de \vr}{\de r}}^{-1},
\end{equation}
where $n_i$, $n_j$, $g_i$, $g_j$ are the number densities and
statistical weights of individual states
$i,j$
that give rise to the studied line, $g_if_{ij}$ is the line oscillator
strength and, $\nu_{ij}$ is the line frequency.
The optically thick lines ($\tau>1$) are important for the radiative
acceleration.
Note that the radiative force due to the optically thick lines does not depend
on number densities of involved atomic states.

Very close to the star, the Sobolev optical depth \eqref{tausob} is large for
many lines both of iron group elements and of lighter elements (like carbon,
nitrogen, etc.) mainly due to 
the
very large wind density. Since iron group elements
have effectively larger number of lines than lighter elements, the former
dominate the radiative force close to the star. Farther out in the wind the line
optical depth
%
\eqref{tausob} is lower (mainly due to decreasing wind density). Since iron
group elements have lower 
abundances
than carbon and nitrogen (i.e.~effectivelly
lower $n_i$), many e.g.~iron lines become optically thin ($\tau<1$) and these
lines do not significantly contribute to the radiative force for higher
wind
velocities. Consequently, the relative contribution of iron group elements to
the radiative force decreases with increasing wind velocity and lighter elements
become more important for the radiative acceleration (see also Vink et
al.~\cite{vikolamet}). This behaviour can be also interpreted as the result of
the fact that lighter elements and iron group elements dominate different parts
of the line-strength distribution function (Puls et al.~\cite{pusle}). In our
case (even for the solar metallicity model) the iron lines become inefficient
already close to the critical point due to a very low wind density.

However, for low metallicities and densities also many lines of lighter
elements that are for higher metallicities optically thick, become
optically thin.
This
leads
to a significant decrease of the mass-loss rate.
In our specific case there is only one optically thick line
(
the
\ion{C}{iii}
resonance line
at
$\lambda=977\,$\AA)
that significantly contributes to the radiative force at
the critical point.
From the contribution of this optically thick line we can
calculate
the value of
the
$\hat\alpha$
parameter
(cf. Puls et al. \cite{pusle}, Eq.\,26)
as $\hat\alpha=0.46$.
Using Eq.~(87) of Puls et al.~(\cite{pusle}), we can roughly estimate
the dependence of the mass-loss rate on the metallicity as $\dot M\sim
Z^{1.2}$.
This
is more consistent
with the decrease of the mass-loss rate with metallicity 
obtained
in our case than
the dependence
$\dmdt\sim Z^{0.67}$ derived by
Krti\v cka (\cite{nlteii}) for 
the
denser wind of
SMC stars.
Note however that the value of $\hat\alpha$ may itself depend on the
metallicity.
In the terms of the 
line-strength distribution
functions
(Puls et al.~\cite{pusle}) the steepening of the dependence of mass-loss
rate on metallicity may be explained as a consequence of the steepening
of the 
line-strength distribution
function for high line-strengths.
Our result
supports the findings of
Kudritzki (\cite{kudmet},
see Fig.~9 therein)
who also found more pronounced dependence of
mass-loss rate on metallicity for
a
%
very
low-metallicity environment. 
%

The calculated wind model is indeed very close to the wind
existence
limit.
As discussed above,
iron lines become inefficient for wind driving and the stellar wind
itself is accelerated mainly due to carbon and nitrogen lines.
CAK critical point is located closer to the stellar
surface
%
as a result of
the decrease of 
a
number of optically thick lines
(c.f.~Babel \cite{babela}, Vink \cite{vintar}).

For the calculation of NLTE models we neglected the multicomponent
effects, however from the values of relative velocity differences
\eqref{xipp}
it is clear, that the stellar wind would decouple relatively close to
the stellar surface.
We have tested 
this possibility
in the three-component
model where the metallic ions are described as one component and
concluded that already in this model the decoupling occurs close to the
stellar surface.

The wind terminal velocity
$v_\infty=1120\kms$
is lower
than in the case of 
a
solar-metallicity model
(cf. Fig. 
\ref{z100q100}).
The lowering of the terminal velocity with respect to the solar
metallicity model is an example for the dependence of the terminal
velocity on metallicity.

The most important parameter that influences the possibility of helium
decoupling is the helium charge.
However, due to 
the
lower wind density, helium is even more ionized
than in the case of the solar metallicity model.

\subsection{Four-component CAK model}

\begin{figure}
\resizebox{\hsize}{!}{\includegraphics{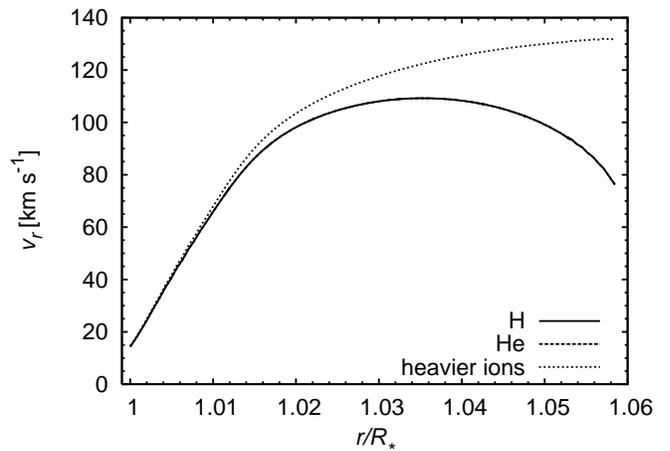}}
\caption{The velocity structure of calculated four-component wind model
with 
%
low metallicity
$Z/Z_\odot=0.1$
and a realistic value of helium charge
$\he z=1$.
Helium velocity
profile
is nearly indistinguishable from
the hydrogen one.
Hydrogen and helium are accelerated at the wind base, however due
to low wind density they decouple
from heavier ions
and may fall back.}
\label{z010q001}
\end{figure}

We have calculated
low-metallicity
four-component wind models with helium as
%
the
fourth component again with the simpler models using 
the
CAK radiative
force.
Although the helium ionization is slightly higher than that
in the
solar metallicity model, we assumed $\he z=1$.
To obtain 
the
lower mass-loss rate inferred from NLTE
calculations we used $k=0.23$.
The other force multipliers are the same as in 
Eq.~\eqref{cakpar}.
It may seem that the adopted value for $k$
was not adequatelly lowered compared to
the value adopted for solar metallicity model
(where $k=0.38$ was used), because $k$ scales with metallicity as
$k\sim Z^{1-\alpha}$ (e.g.~Puls et al.~\cite{pusle}).
However, 
the value $k=0.23$
is appropriate for a four-component wind model of a star with
low metallicity (i.e.~$Z/Z_\odot=0.1$). Because the radiative force
in the four-component models
is
calculated using 
metal
density, 
the decrease of wind metallicity causes lower
radiative force and lower
mass-loss rate.
Consequently, with decreasing metallicity the value of $k$ in a
four-component models varies only slightly.
%

The results 
from
the calculated four-component wind model can be seen in
Fig.~\ref{z010q001}.
Apparently, the friction between metallic ions and the passive component
is low and is not able to accelerate the passive component to velocities
higher than the escape velocity.
Consequently, the wind components decouple and the passive component may
start to fall back to the stellar surface.
However, hydrogen and helium are still coupled together.

From the first point of view it could be expected that there would be a
very high velocity gradient of heavier elements downstream 
from
the point where
the decoupling occurs. However, this is not the case, the velocity increase of
heavier elements is only moderate after the decoupling. The reason 
for
this
behaviour is the same as that which causes non-decoupling described for the
two-component flow by Krti\v cka \& Kub\'at 
(\cite{kk}; see the arguments in the Chapter 3.2 therein).
According to the momentum equation of heavier ions, the radiative force is
approximately equal to the frictional force. Since there is a deficit of the
frictional force in the region where decoupling starts to be effective, the
radiative force is lower there. Consequently, due to the dependence of the
radiative force on the velocity gradient in the Sobolev approximation, the
velocity gradient of heavier ions decreases. Because the velocity of 
the
passive
component (hydrogen and helium) is lower than the escape 
velocity, 
this component
decelerates and may fall back onto the stellar surface
(Porter \& Skouza \cite{reac}).
The wind in this region is likely unstable for the perturbations of
a
multicomponent flow (Owocki \& Puls \cite{op}, Krti\v cka \& Kub\'at
\cite{kkiii}).

The derived terminal velocity 
for the
absorbing ions is very low in this case,
roughly $130\kms$. Consequently, due to low
wind
%
velocities
there would be basically no X-rays emitted in this case by the
possible wind shocks (provided that
the
absorbing ion component does not accelerate
%
as a result of other instabilities).
This model corresponds to models discussed by
Groote \& Schmitt (\cite{grsc}), who argued that the possible explanation of missing periodic X-rays
comming from \sigorie\ may be connected with too low wind velocities.

\section{Discussion}

Although we have used the best models available to us, there are many
simplifications that may slightly modify the results obtained 
(see also
KKII,
%
Krti\v cka \cite{nlteii}).
%

\paragraph{Mass-loss rate:}
%
The
mass-loss rate of \sigorie\ may be actually lower than that obtained
from NLTE wind models, similarly as 
deduced from observations of
O stars with low luminosities (e.g.~Bouret et al. \cite{bourak}, Martins
et al. \cite{martin}).
%
To obtain helium decoupling for \sigorie\ due to 
a
low mass-loss rate,
mass-loss rates of order $10^{-13}\,\text{M}_\odot\,\text{year}^{-1}$
are necessary.
However, in such a case also hydrogen decouples from the flow
(e.g.~KKII).
Therefore, we conclude that the change of the wind mass-loss rate likely
does not influence our basic results.

Helium-rich stars could also eventually have a purely metallic wind, as proposed
by Babel~(\cite{babelb}). However, it is not clear how this could lead to helium
chemical peculiarity.
%

\paragraph{Clumping:}
Winds of OB stars may be clumped (see Martins et al.~(\cite{martin2})
or Fullerton et al.~(\cite{fuj})
for a recent 
observational support of the existence of
wind clumping in O stars).
It is not clear how this effect influences the multicomponent flow.
Whereas the physical state of the interclump medium may significantly
change due to the processes connected with the decoupling,
it is possible that this effect is not very significant
inside
the clumps. This is caused by the fact that
the frictional force depends on the wind density as $\rho^2$, but
on the other hand the mass-loss rate 
derived from 
observations
may be overestimated
due to the clumping.
Possibly, both effects may
mutually
cancel and the net effect of wind clumping
may be marginal.
This picture however significantly depends on the physical properties of
clumps and of surrounding media.
%

\paragraph{Magnetic field:}
We have also neglected the influence of the magnetic field although it
is known that \sigorie\ has a strong one.
The self-consistent treatment of this problem is very complicated and
involves the solution of magnetohydrodynamic equations
(e.g.,
ud-Doula \& Owocki \cite{udo}).
It
should
be noted that for such strong field as that of \sigorie\ this has not
yet been done, even in the one-component approximation, due to extremely
low Courant time 
steps.
An approximate treatment can be done 
by
solving the hydrodynamical
equations along corresponding field lines 
(e.g.~Mestel
\cite{mes2}).
Since this approach applies merely a geometric projection of
corresponding terms in hydrodynamical equations, it is not likely that
this would change the possibility of helium decoupling significantly.

Although MacGregor (\cite{mcg}) showed that strong magnetic field
may significantly influence the wind terminal velocity, 
Owocki \& ud--Doula (\cite{owud}) concluded that the change of the terminal
velocity is only moderate.
The latter
argument
supports
our basic results
to be
valid also in the case of
strong magnetic field.

The magnetic field, however,
may
also change the frictional coefficient due to the particle gyration
induced by the magnetic field or may also
alter
the velocity distribution function of individual species. 
The modification of frictional coefficient due to the magnetic field
could explain
the observational finding
of Groote \& Hunger (\cite{grhu})
that the helium-enrichment
in
\sigorie\
is found where the windflow in the non-magnetic case will be mostly
parallel to the magnetic field lines of this star.
Similar findings for other helium strong stars were reported by Groote
(\cite{gro}).

The importance of the magnetic field for the collision frequency
of helium atoms
%
is given by the value of
the
product of 
the
mean collision time 
between
helium ions 
and
%
protons
$\hevo\tau$ and gyration frequency of helium ions $\he\omega$
(Mestel \cite{meskni}, page 18 therein; Braginskij \cite{brag},
Imazu \cite{gerisiteimazu}).
The mean collision time $\hevo\tau$ is given by
\begin{equation}
\label{tauro}
\hevo\tau^{-1}=\frac{2}{3\sqrt\pi}\frac{\hevo K}{\hevo\alpha\he\rho},
\end{equation}
where 
the
frictional coefficient $\hevo K$ is given by Eq.~\eqref{trekon},
and $\hevo\alpha$ is given by Eq.~\eqref{alrov}.
The gyration frequency is
\begin{equation}
\label{omegaro}
\he\omega=\frac{\he q B}{\he mc},
\end{equation}
where $B$ is the magnetic field intensity.
For collision-dominated plasmas
$\hevo\tau\he\omega\ll1$ the influence of 
the
magnetic field on the momentum
transport between individual plasma components can be neglected.
In the opposite case $\hevo\tau\he\omega\gg1$
a
%
strong anisotropy of microscopic plasma properties 
may occur
(in the directions
parallel and perpendicular to the magnetic field).
%
However, the frictional coefficients parallel to the magnetic field
remain the same as in the case without magnetic field.
Consequently, the change of frictional coefficients due to the magnetic
field is not important in the case when magnetic energy density
dominates over the gas kinetic energy
density,
i.e.~in the case when plasma moves along magnetic field lines.

Application of Eqs.~\ref{tauro},~\ref{omegaro} to our
derived wind model has shown that
for typical magnetic field with intensities of
the order of
%
$10^4\,$G,
which
corresponds to
%
the surface magnetic field of \sigorie\ derived by
Landstreet \& Borra (\cite{labor}),
%
in the outer parts of \sigorie\ wind the inequality
$\hevo\tau\he\omega\gg1$ holds.
This inequality holds
even for
lower magnetic field intensities of the
order of $10^2\,$G (the lower value of the magnetic field intensity roughly
takes into the account its decrease with radius and its latitudinal
variations).
Since wind flows along magnetic field lines due to the strong wind
confinement by the magnetic field, the above mentioned variation of
frictional coefficient with magnetic field is likely not very important
in our case.
%

\paragraph{Ionization:}
Our results are sensitive to the ionization state of helium atoms.
However, 
because
derived helium charge is
by
two orders of magnitude higher than
the value
necessary for helium decoupling, it seems that our results are also
robust with respect of errors in the calculation
of statistical equilibrium equations
(note that helium is dominated by its first ionization stage already in the stellar atmosphere).

Even in the case when doubly ionized helium comprises a significant
fraction of all helium atoms, some of them remain neutral
(Fig.~\ref{helna}).
This means that each helium atom for some (however very small) fraction
of time is neutral.
Recall that in the case of \sigorie\ some wind material may remain
trapped in the minima of gravitational and centrifugal potential along
magnetic field lines (Townsend \& Owocki \cite{towo}).
The 
non-zero
fraction of time spend by each helium atom in the neutral
state may then induce drift of helium atom across 
magnetic field
lines.
Let us at least approximatively derive the magnitude of such 
an
effect.
Recall that the 
typical time of neutral helium ionization is
$\tau_\text{ioniz}$
(see Eq.~%
\eqref{tauioniz})
and 
%
%
the typical time of neutral helium recombination is
introduced similarly as
$\tau_\text{rec}$.
During the time $\tau_\text{ioniz}$ the helium atom is not supported by
the magnetic field
(since it is neutral)
and
its
absolute value of the radial velocity increases
by
(neglecting the centrifugal force)
$g\tau_\text{ioniz}$ where $g=GM/r^2$.
During the time $\tau_\text{rec}$ the helium atom is supported by the
magnetic field and has zero radial velocity.
Assuming that helium is mostly ionized,
$\tau_\text{rec}\gg\tau_\text{ioniz}$,
%
this 
means
that 
the
average velocity is roughly
$
g\tau_\text{ioniz}^2/(\tau_\text{rec}+\tau_\text{ioniz}) \approx
g\tau_\text{ioniz}^2/\tau_\text{rec}$. The typical time
necessary to travel the distance $R_*$ is
$\tau_\text{diff}=R_*^3\tau_\text{rec}/\zav{GM\tau_\text{ioniz}^2}$.
For
%
typical values for the outer wind parts $\tau_\text{rec}\approx 10^5\,$s
and
$\tau_\text{ioniz}\approx 1\,$s and the typical diffusion time
$\tau_\text{diff}
\approx10^5$~years
is too large compared to the typical time for emptying of 
the
magnetosphere,
which is of 
the
order of 
a
hundred years (Townsend \& Owocki \cite{towo}).
Moreover, the drift may occur in 
outward direction,
%
since the total force acting on neutral helium 
atoms
(i.e.~the
centrifugal force plus the gravitational force) may be directed outwards
and in such a case no fall back 
will 
occur.
%
Note that the rotation of the magnetosphere containing charged particles
may be roughly approximated as a rigid body rotation, at least near 
to
the
stellar surface.
%


\section{Conclusions}

We have studied whether the helium decoupling can occur in the stellar
wind of \sigorie\ and, consequently, we tested whether 
this process
may consistently explain enhanced helium
abundance in the atmospheres of Bp stars (as proposed by Hunger \&
Groote \cite{HuGr}).

For this purpose we first calculated 
an
NLTE wind model 
for
\sigorie\ and
derived its mass-loss rate, terminal velocity and helium ionization.
These parameters were consequently used for the calculation of
four-component wind models to test whether helium decoupling may occur.
As a by-product of our study we have found that the frictional heating
due to collisions between individual heavier elements (for example
nitrogen) and 
the
passive component (hydrogen and helium) is important in
the outer wind regions of \sigorie.

Our four-component wind models showed that for realistic values of
helium charge the decoupling of helium from the stellar wind of
\sigorie\ is unlikely.
Helium decoupling is possible only for very low helium charge, much
lower than 
derived from NLTE equations.
We discussed the possibilities that may change this result, however, we
were not able to find any consistent model that could lead to 
helium decoupling in the wind 
under such conditions.
%
%

The main result of our paper,
%
that for realistic values of helium
charge helium will not decouple from the wind flow for parameters
suitable for a star 
like
\sigorie,
is
in contradiction to the result of Hunger \& Groote (\cite{HuGr}) who
obtained 
the
opposite behaviour.
The reason for this is that helium is strongly coupled to
\emph{hydrogen}
and this coupling cannot be
neglected in 
model calculations.

We have shown that for lower metallicities the wind mass-loss rate is
significantly lower.
This is caused by a significant decrease of 
the
number of optically thick lines.
For such low mass-loss rates the decoupling of wind components and
a
subsequent fall back of the passive component may occur.
However, helium and hydrogen remain coupled in this case, consequently
this does not
explain
the chemical peculiarity.

We studied the multicomponent flow analytically and derived a simple
condition for helium decoupling in the stellar wind.
This analysis supports our basic result that helium decoupling is
possible only for a very low helium charge (i.e.~
when
helium is basically neutral).

We have also studied the possibility of decoupling of neutral helium.
This could be an interesting way of helium decoupling (at least
partially), 
since the cross-section of collisions of protons with neutral
helium is much lower than the effective
cross-section for collisions between
charged particles.
However, the neutral helium atoms do not remain
%
neutral for sufficiently 
long
time since 
the
characteristic helium
ionization time is much shorter than
the characteristic time of change
in
flow properties $\tau_\text{flow}$.

As observations 
seem to
favour the hypothesis of helium decoupling 
(Groote \& Hunger \cite{grhu}, Hunger \& Groote \cite{HuGr}) our result
leaves us in the very unpleasant situation that we do not have any
self-consistent explanation for 
the
chemical peculiarity of He-strong stars.
While the scenario proposed by Hunger \& Groote (\cite{HuGr})
requires
artificial lowering of the helium charge, models for helium
diffusion moderated by the stellar wind (Vauclair \cite{vauche},
Michaud et al. \cite{mihelpek})
require artificial lowering of the wind mass-loss rate.
The only remaining 
%
influence on the helium decoupling,
which
is very difficult to calculate
is that of a strong magnetic field. 
Nevertheless, it seems clear that decoupling of helium in normal
(non-magnetic) stars with comparable stellar parameters can be excluded
as a mechanism of creating the helium overabundance in the stellar
atmosphere.
Moreover, we 
were
not able to find any self-consistent mechanism that would
enable the decoupling in the case of magnetic stars.

\appendix
\section{Approximate description of H-He-i flow}
\label{dvprib}

In order to obtain an approximate description of a well-coupled
multicomponent flow of hydrogen, helium and heavier ions%
\footnote{Note that in this Appendix all heavier ions are described as
%
one component.}
(neglecting electrons) we can write 
the
momentum equations of hydrogen
(H)
and helium
($\alpha$)
as (Burgers \cite{burgers}; neglecting gas-pressure term and
polarization electric field)
\begin{subequations}
\label{zahyb}
\begin{align}
\vo\vr\frac{\de\vo\vr}{\de r}=&-g-
\frac{\vohe K}{\vo\rho} 
\Gch
(\vohe x)\frac{\vohe x}{|\vohe x|}+
\frac{\voio K}{\vo\rho} 
\Gch
(\voio x)\frac{\voio x}{|\voio x|},\\*
\he\vr\frac{\de\he\vr}{\de r}=&-g+
\frac{\vohe K}{\he\rho} 
\Gch
(\vohe x)\frac{\vohe x}{|\vohe x|}+
\frac{\heio K}{\he\rho} 
\Gch
(\heio x)\frac{\heio x}{|\heio x|}.
\end{align}
\end{subequations}
%
We assume that the flow is well-coupled,
$\vo\vr\frac{\de\vo\vr}{\de r}\approx\he\vr\frac{\de\he\vr}{\de r}
\equiv\vr\frac{\de\vr}{\de r}$, and we
%
use only the linear term of
the Taylor expansion of the Chandrasekhar function
$
\Gch
(x)\approx\frac{2|x|}{3\sqrt{\pi}}$.
Then
we can rewrite Eqs.~(\ref{zahyb}) as
\begin{subequations}
\begin{align}
\label{prihybh}
\vr\frac{\de\vr}{\de r}=&-g-
\frac{\vohe k}{\vo\rho}\frac{\Delta\vohe\vr}{\vohe\alpha}+
\frac{\voio k}{\vo\rho}\frac{\Delta\voio\vr}{\voio\alpha},\\*
\label{prihybhe}
\vr\frac{\de\vr}{\de r}=&-g+
\frac{\vohe k}{\he\rho}\frac{\Delta\vohe\vr}{\vohe\alpha}+
\frac{\heio k}{\he\rho}\frac{\Delta\voio\vr+\Delta\vohe\vr}{\heio\alpha},
\end{align}
\end{subequations}
where
we used $x_{ab} = \Delta\vr_{ab}/ \alpha_{ab}$. Here
$k_{ab}=2K_{ab}/\zav{3\sqrt{\pi}}$ and the velocity difference
$\Delta\vr_{ab}=\vr_a-\vr_b$
for all $a$, $b$, which stand for $\mathrm{H}$, $\alpha$, and
$\mathrm{i}$.
Subtracting Eqs.~(\ref{prihybhe}) and (\ref{prihybh}) we obtain
\begin{multline}
\Delta\voio\vr\zav{\frac{\voio k}{\vo\rho\voio\alpha}-
                   \frac{\heio k}{\he\rho\heio\alpha}}
		= \\*
\Delta\vohe\vr\zav{\frac{\vohe k}{\he\rho\vohe\alpha}+
                   \frac{\heio k}{\he\rho\heio\alpha}+
		   \frac{\vohe k}{\vo\rho\vohe\alpha}}.
\end{multline}
Due to the very low number density of heavier ions $\io n$, we can
neglect the second right hand side
term and derive (assuming that the temperatures of all wind components
are equal;
note that the similar term on the left hand side can not be neglected because
all 
terms
on the left hand side are of the same order)
\begin{equation}
\label{dvoiovohe}
\Delta\voio\vr\frac{\io n}{\he n}
  \frac{\io z^2}{\he z^2}\frac{\vohe\alpha}{\voio\alpha}
  \zav{1-\frac{\vo m\voio\alpha\he z^2}{\he m\heio\alpha\vo z^2}}
  \zav{1+\frac{\vo\rho}{\he\rho}}^{-1}
=\Delta\vohe\vr.
\end{equation}
The sign of $\Delta\vohe\vr$ depends on the sign of the first bracket in
Eq.~(\ref{dvoiovohe}). For
\begin{equation}
1-\frac{\vo m\voio\alpha\he z^2}{\he m\heio\alpha\vo z^2}>0,
\end{equation}
i.e.~for 
a
low helium charge, the hydrogen velocity is higher than the helium
velocity. Because $\io n\ll\he n$ we derive $\Delta\voio\vr\gg\Delta\vohe\vr$
for a well coupled flow and realistic values of 
the
other parameters. 
Helium
decoupling may occur in the case $\Delta\vohe\vr\gg\Delta\voio\vr$. Apparently,
this may occur for very low helium charge $\he z$, as was demonstrated in the
paper.

Finally, let us compare the frictional force acting on helium due to 
heavier
ions and
hydrogen in the case of 
a
well-coupled flow. The frictional acceleration 
after
Eq.~(\ref{rovhyb}), assuming $\io\vr>\vo\vr>\he\vr$ and
$\Delta\voio\vr\gg\Delta\vohe\vr$,
is
\begin{multline}
\label{hefric}
\frac{1}{\he\rho}\heio K 
\Gch
(\heio x)+\frac{1}{\he\rho}\vohe K 
\Gch
(\vohe x)\approx
\\* \frac{\heio k}{\he\rho}\frac{\Delta\voio\vr}{\heio\alpha}
\zav{1+\frac{\vohe k}{\heio k}\frac{\heio\alpha}{\vohe\alpha}
     \frac{\Delta\vohe\vr}{\Delta\voio\vr}}.
\end{multline}
Inserting Eq.~(\ref{dvoiovohe}) for the fraction of velocity differences and
performing just an order of magnitude estimate we may assume
$\heio\alpha\approx\vohe\alpha$, $\io z\approx\he z$ and
$\vohe\alpha\approx\voio\alpha$ and 
%
conclude that both terms in
\begin{equation}
1+\frac{\vohe k}{\heio k}\frac{\heio\alpha}{\vohe\alpha}
     \frac{\Delta\vohe\vr}{\Delta\voio\vr}
\end{equation}
are of the same order. This means that the contribution of heavier ions and
hydrogen to the helium acceleration are of the same order in the case of a
well-coupled flow (e.g. for O supergiants). However, in the case when heavier
ions are not able to effectively accelerate helium the friction with hydrogen
dominates.

\begin{acknowledgements}
We would like to thank to Dr.  Marian Karlick\'y for the discussion of
the influence of the magnetic field on the frictional coefficients,
and to Dr. Joachim Puls
for his valuable comments on the manuscript.
This research has made use of NASA's Astrophysics Data System and the
SIMBAD database, operated at CDS, Strasbourg, France.
This work was supported by grants GA \v{C}R 205/03/D020, 205/04/1267.
The Astronomical Institute Ond\v{r}ejov is supported
by project
AV0\,Z10030501.
\end{acknowledgements}

\newcommand{\actob}{in Active OB-Stars: Laboratories for Stellar
	\& Circumstellar Physics, S. \v{S}tefl, S. P. Owocki, \& A.~T.
	Okazaki eds., ASP Conf. Ser., submitted}
%


\end{document}